\documentclass[conference]{IEEEtran}
\IEEEoverridecommandlockouts
\usepackage{cite}
\usepackage{amsmath,amssymb,amsfonts}
\usepackage{algorithmic}
\usepackage{graphicx}
\usepackage{textcomp}
\usepackage{xcolor}
\def\BibTeX{{\rm B\kern-.05em{\sc i\kern-.025em b}\kern-.08em
    T\kern-.1667em\lower.7ex\hbox{E}\kern-.125emX}}
    
    \allowdisplaybreaks
\begin{document}

\title{Analyzing Power Quality Implications of High Level Charging Rates of Electric Vehicle Within Distribution Networks}

\author{
\IEEEauthorblockN{1\textsuperscript{st} Arash Farokhi Soofi}
\IEEEauthorblockA{\textit{Electrical and Computer Engineering} \\
\textit{University of California, San Diego}\\
San Diego, USA 92161 \\
afarokhi@ucsd.edu}
\and
\IEEEauthorblockN{2\textsuperscript{nd} Reza Bayani}
\IEEEauthorblockA{\textit{Electrical and Computer Engineering} \\
\textit{University of California, San Diego}\\
San Diego, USA 92161  \\
rbayani@ucsd.edu}
\and
\IEEEauthorblockN{3\textsuperscript{nd} Saeed D. Manshadi}
\IEEEauthorblockA{\textit{Electrical and Computer Engineering} \\
\textit{San Diego State University}\\
San Diego, USA 92182 \\
smanshadi@sdsu.edu}}

\maketitle
\begin{abstract}
    This paper investigates the impact of the charging level of high penetration level of Electric Vehicles (EVs) on the power quality of the electricity distribution network. The EV owners tend to charge their EVs as fast as possible. The charging levels of EVs within the distribution network affect the voltage profile of buses of the network. In this paper, an exact Second-Order Cone Programming (SOCP) formulation of the full AC optimal power flow (ACOPF) problem of the distribution network is presented. The network includes solar generation units and EVs as Distributed Energy Resources (DERs).  Different charging levels are considered to analyze the impact of EVs on the distribution network. The performance of the proposed model is illustrated for the modified IEEE-33 bus system for different charging levels for EVs. Besides, the impact of available solar power and battery degradation cost of EVs on the distribution network is investigated. It is illustrated that how EV charging will cause voltage deviation challenges for the distribution network.
\end{abstract}

\section*{Nomenclature}
\subsection*{Parameters}
\noindent \begin{tabular}{ l p{7cm} }
$A_{s}^{t}$ & Available power of solar generation unit $s$ at time $t$ \\
$B_{(.)},G_{(.)}$  & Elements of susceptance and conductance matrices\\
$C_D$ & Battery degradation cost \\
$C_E$ & Time-of-use price of electricity at time $t$ \\
$\overline{E_{e}}$, $\underline{E_{e}}$  & Maximum and Minimum energy of EV $e$\\
$P_d^t$,$Q_d^t$ & Real and reactive power of load $d$ at time $t$\\
$P_{e,t}^{tr}$ & Traveling power consumption of EV $e$ at time $t$ \\
$R_d^t$, $R_c^t$ & Ratio of dis/charging of EVs to the total EVs at time $t$ \\
$\overline{S}_{ij}$ & Maximum apparent power flow of distribution line \\
$\overline{V}_i,\underline{V}_i$  & Maximum/Minimum voltage magnitude at bus $i$\\
$\gamma_e^c,\gamma_e^d$ & Charging and discharging efficiency of EV $e$ 
\end{tabular}

\subsection*{Sets}
\noindent \begin{tabular}{ l p{7cm} }
$\mathcal{D}_i$ & Set of load connected to bus $i$ \\
$\mathcal{E}_i$ & Set of the electric vehicle connected to bus $i$ \\
$\mathcal{G} $ & Set of all grids connected to the network \\
$\mathcal{G}_{i} $ & Set of the grid connected bus $i$\\
$\mathcal{S}$ & Set of all solar generation units \\
$\mathcal{S}_i$ & Set of solar generation unit connected to bus $i$ \\
$\mathcal{T} $ & Set of time horizon 
\end{tabular}

\subsection*{Variables}
\noindent \begin{tabular}{ l p{7.1cm} }
$e_i$, $f_i$ & Real and Imaginary part of voltage phasor of bus $i$\\
$E_{e,t}$ & Energy of EV $e$ at time $t$\\
$I_{e,t}^{c,j}$& Binary variable representing charging level of EV $e$ at time $t$\\
$P_s^t$ & Real power dispatch of solar generation unit $s$ at time $t$\\
$P_{e,t}^{c}$  & Real charging power of EV $e$ at time $t$ \\
$P_{e,t}^{c,j}$  & Real charging power of level $j$ of EV $e$ at time $t$ \\
$P_g^t,Q_g^t$  & Real/Reactive power flow between grid $g$ and the distribution network at time $t$\\
$P_{ij}^t,Q_{ij}^t$ & Real/Reactive power flow between bus $i$ and $j$ at time $t$\\
$c,s$ & Lifting operator terms of SOCP relaxation method\\
$V_{i}^t$ & Voltage phasor of bus $i$ at time $t$ 
\end{tabular}

\section{Introduction}
Utilizing and developing EVs is an effective method to mitigate CO2 pollution and decrease dependence on fossil fuel by many countries \cite{xia2010evaluation}, \cite{jian2012regulated}. Thus, a notable number of EV charging facilities within the distribution network are required to support the push toward transportation electrification. One major concern of EV owners is the waiting time to charge which is in turn dependent on the charging level of EVs. EV owners prefer to charge their vehicles as soon as possible. In recent years, different types of EV batteries and charging stations with different charging levels are introduced \cite{veneri2012charging}, \cite{falvo2014ev}. The impact of EV charging on the grid depends on the charging power delivered to the charging station. Investigating the impacts of high EV charging levels is crucial for utilities to adopt the appropriate strategies to enhance the power quality of the distribution network. There are various methods and apparatuses in the literature to enhance the power quality of the distribution network \cite{jafari2020new,mazaheri2021investigating,babaei2021data,tran2018solar,taghavirashidizadeh2020genetic,farhoodnea2013power}. The analysis of a single DER and its impact on the system is investigated in \cite{delimustafic2011model}. 
However, the power quality implications (e.g. voltage) of high-level EV chargers in a distribution network with a high penetration of EVs are not yet investigated in the literature.
\\
This paper aims to investigate the impact of utilizing high-level EV chargers on the power quality of the distribution system. The full AC Optimal Power Flow (ACOPF) problem of the distribution network with Photovoltaic (PV) and EV charging stations are employed to investigate this impact. 
This paper aims to find the answers to these questions: \textit{What is the impact of utilizing fast-charging EV chargers on the electricity distribution network? What is the impact of increasing the size of solar generation units and available solar generation on the voltage profile of buses when fast-charging EV chargers are utilized? What is the impact of battery degradation cost on the power quality of the distribution network at different buses?}

\section{Problem Formulation and Solution Method}
The ACOPF optimization problem of the distribution network with solar generation units and fixed power model of EVs with different charging levels is presented in this section. Then, due to the non-convexity of the ACOPF problem, the SOCP lifting variables are presented to reformulate the ACOPF problem formulation as a convexified optimization problem.
\subsection{Objective Function}
The objective function of the ACOPF problem of the distribution network with DERs is presented in \eqref{SOCP_obj}. The ACOPF optimization problem of the distribution networks minimizes the operation cost of the distribution network and DERs. The first part of the objective function represents the operational cost of the distribution network which is the summation of the main power grid injection to the distribution network multiplied by the time-of-use price over $24$ hours. The second part of the objective function represents the battery degradation cost of EVs over $24$ hours based on their charging level. The cost of battery degradation increases when the charging level of EVs increases. The total battery degradation cost is the summation of charging power of different charging levels multiplied by the degradation cost of each charging level.

 \begin{equation} 
 \text{min}\sum_{g\in  \mathcal{G}}^{ }\sum_{t \in \mathcal{T}}^{ }C_E^tP_{g}^t+\sum_{e\in  \mathcal{E}}^{ }\sum_{t \in \mathcal{T}}^{} \sum_{j \in \mathcal{J}}^{} C_D{P_{e,t}^{c,j}}^2\label{SOCP_obj}
\end{equation} 

\subsection{EV and PV Systems Model}
The model of EVs with different charging levels is presented in \eqref{EV}. The upper and lower limits of each charging level of EVs is presented in \eqref{SOCP_mu_pe_lev}. Inequality \eqref{bin_level} shows that the mutually exclusiveness of charging level of EV $e$. Thus, only one of the charging power levels of EV $e$ can be nonzero at time $t$. The summation of all charging power levels of EV $e$ at time $t$ is equal to the charging power of EV $e$ at $t$ as presented in \eqref{SOCP_lambda_pe}. Constraint \eqref{SOCP_mu_Ee} represents the energy limit of EV $e$ at time $t$. The energy balance equation of EV $e$ at time $t$ is presented in \eqref{SOCP_lambda_Eet} and \eqref{SOCP_lambda_Ee1}. The power generated by PV $s$ at time $t$ is less than or equal to the available solar power for PV $s$ at time $t$ as shown in \eqref{SOCP_mu_s}.
\begin{subequations}\label{EV}
\begin{alignat}{2}
&0\leq P_{e,t}^{c,j}\leq I_{e,t}^{c,j}\overline{P_{e}^{c,j}}R_c^t\label{SOCP_mu_pe_lev}\\
&\sum_{j \in \mathcal{J}}^{}I_{e,t}^{c,j}\leq 1 \label{bin_level}\\
&\sum_{j \in \mathcal{J}}^{} P_{e,t}^{c,j}= P_{e,t}^{c}\label{SOCP_lambda_pe}\\
& \underline{E_{e}}\leq E_{e,t}\leq \overline{E_{e}}\label{SOCP_mu_Ee}\\
&E_{e,t}=E_{e,t-1}-(\frac{P_{e,t}^{tr}R_d^t}{\gamma_e^d}-\gamma_e^cP_{e,t}^c) \hspace{0.2cm}  \forall{t \in \mathcal{T} \setminus 1}\label{SOCP_lambda_Eet}\\
&E_{e,t}=E_{e,t-1+|T|}-(\frac{P_{e,t}^{tr}R_d^t}{\gamma_e^d}-\gamma_e^cP_{e,t}^c)\hspace{0.2cm}    \forall{t=1}\label{SOCP_lambda_Ee1}\\
&0\leq P_{s}^t\leq A_{s}^t\label{SOCP_mu_s}
\end{alignat}
\end{subequations} 
 \subsection{Nodal Balance and Power Flow Constraints}
 The AC real and reactive power flow equations are presented in \eqref{AC_p} and \eqref{AC_q}, respectively. The real and reactive nodal balance of distribution network is presented in \eqref{AC_lambda_i_p} and \eqref{AC_lambda_i_q}, respectively. The line limit of the distribution network is shown in \eqref{AC_lim}.
 \begin{subequations} \label{AC_PQ}
\begin{alignat}{2}
&\left\{\begin{matrix}
\hspace{-1.9cm}P_{ij}^t=-G_{ij}({e_i^t}^2+{f_i^t}^2)+
\\ 
 G_{ij}(e_i^te_j^t+f_i^tf_j^t)-B_{ij}(e_i^tf_j^t-e_j^tf_i^t)
\end{matrix}\right.\hspace{0.3cm}\forall{(i,j)} \in \mathcal{L} \label{AC_p}\\
&\left\{\begin{matrix}
\hspace{-1.9cm}Q_{ij}^t=B_{ij}({e_i^t}^2+{f_i^t}^2)-
\\ 
B_{ij}(e_i^te_j^t+f_i^tf_j^t)-G_{ij}(e_i^tf_j^t-e_j^tf_i^t)
\end{matrix}\right.\hspace{0.2cm}\forall{(i,j)} \in \mathcal{L}
\label{AC_q}\\ 
&\sum_{s \in \mathcal{S}_i}^{} P_s^t+\sum_{g \in \mathcal{G}_i } P_g^t=\sum_{d \in \mathcal{D}_i}^{}P_d^t+(G_{ii}+\sum_{j\in \mathcal{\delta}_i}^{}G_{ij})({e_i^t}^2+{f_i^t}^2)\nonumber
\\
&\hspace{1cm}+\sum_{j\in \mathcal{\delta}_i}P_{ij}^t+\sum_{e\in\mathcal{E}_i}^{}P_{e,t}^c
\hspace{0.8cm}\forall{i \in \mathcal{N}},\forall{(i,j)} \in \mathcal{L}\label{AC_lambda_i_p}\\
&\sum_{g \in \mathcal{G}_i}^{}Q_g^t-\sum_{d \in \mathcal{D}_i}^{}Q_d^t=-(B_{ii}+\sum_{j\in \mathcal{\delta}_i}^{}B_{ij})({e_i^t}^2+{f_i^t}^2)+\sum_{j\in \mathcal{\delta}_i}Q_{ij}^t\nonumber\\
&\hspace{5cm}\forall{i \in \mathcal{N}},\forall{(i,j)} \in \mathcal{L}\label{AC_lambda_i_q}\\
&\sqrt{(P_{ij}^t)^2+(Q_{ij}^t)^2}\leq  \overline{S}_{ij}\hspace{2.6cm}\forall{(i,j)} \in \mathcal{L}\label{AC_lim}
\end{alignat}
\end{subequations} 
 The source of non-linearity of the ACOPF optimization problem is power flow and nodal balance equations. Thus, a set of SOCP lifting variables is presented in \eqref{lifting} to relax the power flow and nodal balance equations. The SOCP relaxed form of the power flow and nodal balance equations of the distribution networks with DERs is presented in \eqref{SOCP_PQ}. It should be noted that the SOCP relaxation method is exact for acyclic distribution networks \cite{Soofi2021NAPS,soofi2020socp}.
\begin{subequations}  \label{lifting}
\begin{alignat}{2}
&c_{ii}^t:={e_i^t}^2+{f_i^t}^2 \label{cc}\\
&c_{ij}^t:=e_i^te_j^t+f_i^tf_j^t=\textbf{Re}\{V_i^t{V_j^t}^*\}\label{cij}\\
&s_{ij}^t:=e_j^tf_i^t-f_j^te_i^t=\textbf{Im}\{V_i^t{V_j^t}^*\}\label{sij}
\end{alignat} 
\end{subequations}
 The relaxed form of the real and reactive power flow equations are presented in \eqref{SOCP_p} and \eqref{SOCP_q}, respectively. The relaxed form of the real and reactive nodal balance of the distribution network is presented in \eqref{SOCP_lambda_i_p} and \eqref{SOCP_lambda_i_q}, respectively. The line limit of the distribution network is the same as the one presented in \eqref{AC_lim}. 
\begin{subequations} \label{SOCP_PQ}
\begin{alignat}{2}
&P_{ij}^t=-G_{ij}c_{ii}^t+G_{ij}c_{ij}^t+B_{ij}s_{ij}^t\hspace{0.3cm}\forall{i \in \mathcal{N}},\forall{(i,j)} \in \mathcal{L}\label{SOCP_p}\\
&Q_{ij}^t=B_{ij}c_{ii}^t-B_{ij}c_{ij}^t+G_{ij}s_{ij}^t\hspace{0.6cm}\forall{i \in \mathcal{N}},\forall{(i,j)} \in \mathcal{L}\label{SOCP_q}\\
&\sum_{s \in \mathcal{S}_i}^{} P_s^t+\sum_{g \in \mathcal{G}_i}^{}P_g^t =\sum_{d \in \mathcal{D}_i}^{}P_d^t+(G_{ii}+\sum_{j\in\delta_i}^{}G_{ij})c_{ii}^t+\nonumber\\
&\hspace{1.5cm}\sum_{j \in \delta_i}^{}P_{ij}^t+\sum_{e\in\mathcal{E}_i}^{}P_{e,t}^c\hspace{0.6cm}\forall{i \in \mathcal{N}},\forall{(i,j)} \in \mathcal{L}\label{SOCP_lambda_i_p}\\
&\sum_{g \in \mathcal{G}_i}^{}Q_g^t-\sum_{d \in \mathcal{D}_i}^{}Q_d^t=-(B_{ii}+\sum_{j\in \delta_i}^{}B_{ij})c_{ii}^t+\sum_{j \in \delta_i}^{}Q_{ij}^t\nonumber\\
&\hspace{4.6cm}\forall{i \in \mathcal{N}},\forall{(i,j)} \in \mathcal{L}\label{SOCP_lambda_i_q}
\end{alignat}
\end{subequations}

\subsection{Voltage and SOC Constraints}
The lower and upper limits of voltage magnitude of bus $i$ at time $t$ are presented in \eqref{SOCP_limv}. The inequality constraints on the left-hand side of the arrow are the AC form of the voltage limit constraints. Leveraging the SOCP lifting variables presented in \eqref{lifting}, the SOCP reformulated form of the voltage limit constraints is presented on the right-hand side of the arrow. The relation between the SOC lifting variables is presented in \eqref{SOCP_symtric} and \eqref{SOC}. 
\begin{subequations}\label{volt}
\begin{alignat}{2}
&{\underline{V}_i}^2\leq  {e_i^t}^2+{f_i^t}^2\leq {\overline{V}_i}^2\overset{\eqref{lifting}}{\rightarrow} {\underline{V}_i}^2\leq  c_{ii}^t\leq {\overline{V}_i}^2\label{SOCP_limv}\\
&c_{ij}^t=c_{ji}^t \hspace{1cm} ,\hspace{1cm} s_{ij}^t=-s_{ji}^t\label{SOCP_symtric}\\
&    \begin{Vmatrix}
2c_{ij}^t
\\2s_{ij}^t
\\ 
c_{ii}^t-c_{jj}^t
\end{Vmatrix}\leq c_{ii}^t+c_{jj}^t\label{SOC}
\end{alignat}
\end{subequations}

\subsection{The Relaxed Problem Formulation}
The SOCP convexified optimization problem of the distribution network with solar generation units and EV stations with different charging levels is presented in \eqref{SOCP_ref}.

\begin{subequations}\label{SOCP_ref}
\begin{alignat}{2}
&\underset{P_g^t,P_{e,t}^{c,j}}{\text{min}}\sum_{g\in  \mathcal{G}}^{ }\sum_{t \in \mathcal{T}}^{ }C_E^tP_{g}^t+\sum_{e\in  \mathcal{E}}^{ }\sum_{t \in \mathcal{T}}^{} \sum_{j \in \mathcal{J}}^{} C_D{P_{e,t}^{c,j}}^2\label{obj_SOC_ref}\\
& \text{s.t.} \hspace{1cm} \eqref{EV}, \eqref{AC_lim}, \eqref{SOCP_PQ}, \eqref{volt}
\end{alignat}
\end{subequations}

\section{Test System}
\begin{figure}[t!]
\centering
\begin{minipage}{.45\textwidth}
  \centering
    \includegraphics[width=\columnwidth]{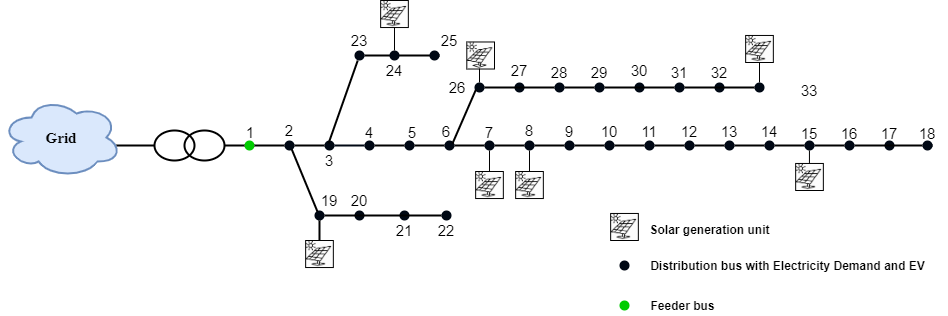}
    \caption{Modified IEEE 33-bus distribution network}
    \label{fig:33-bus}
\end{minipage}
\hspace{1cm}
\begin{minipage}{.45\textwidth}
  \centering
  \includegraphics[width=\columnwidth]{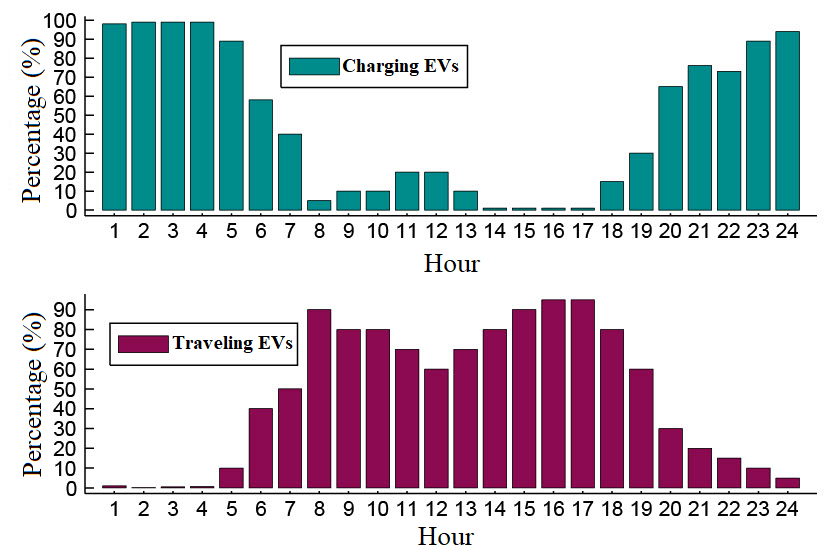}
  \caption{Daily distribution of traveling and charging EVs}
  \label{fig:flag}
\end{minipage}
\end{figure}

The topology of the modified IEEE 33-bus system is presented in Fig. \ref{fig:33-bus}. The test system consists of 33 buses, 32 branches, 7 solar generation units, 32 loads, and 32 EV charging points. The base net demand and solar generation trends are set according to the normalized hourly demand of California ISO on August 18, 2020. Each EV could have one of these three conditions: traveling, charging, or parking without any connection to the distribution network. The percentage of EVs which are traveling and connected to the distribution network at each time of day is presented in Fig. \ref{fig:flag}. The number of EVs connected to each bus connected with EV charging point is $10$. 

\section{Results}
\subsection{The Impact of Different EV Charging Levels on the Operation of Distribution Network}
\begin{figure}[h!]
\centering
  \centering
  \includegraphics[width=\columnwidth]{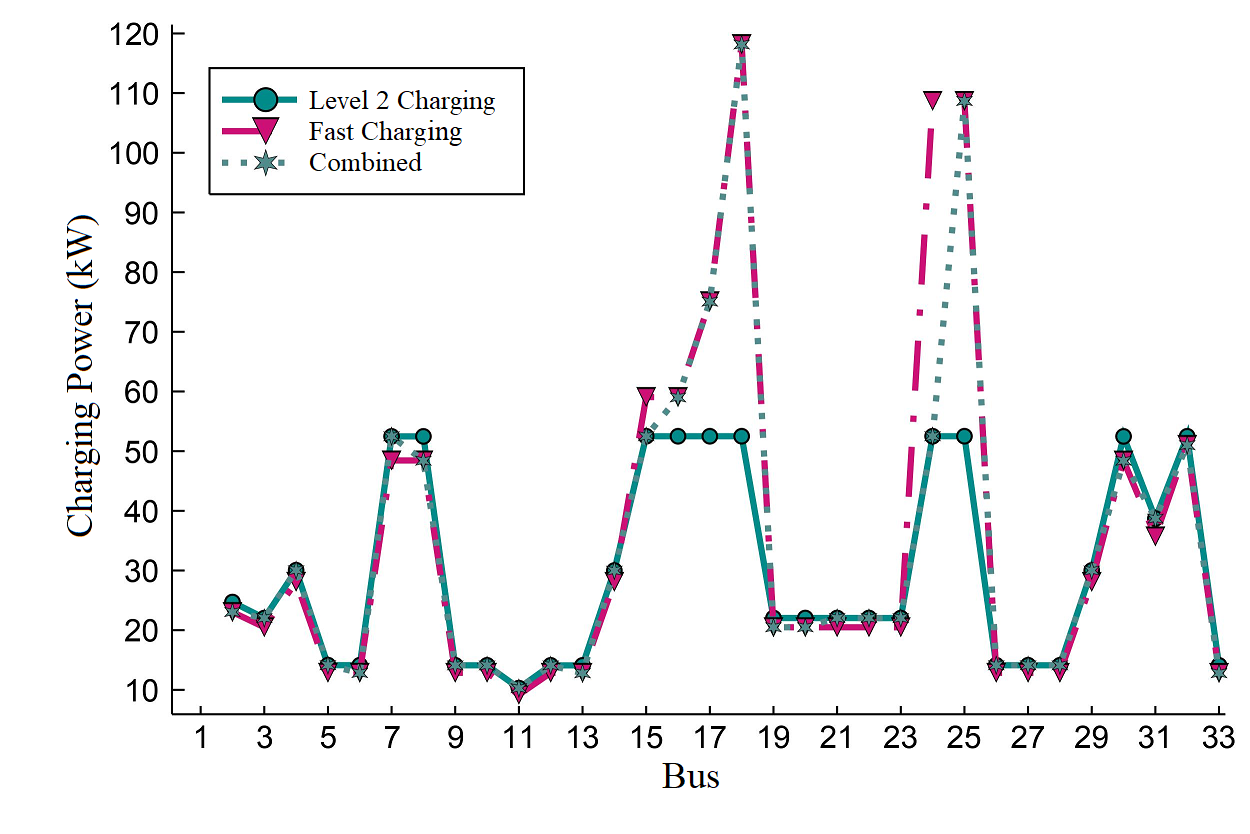}
  \caption{Comparing EV charging power of various scenarios at $6$ p.m.}
  \label{fig:power_time_18}
\end{figure}

\begin{figure}[h!]
 \centering
    \includegraphics[width=\columnwidth]{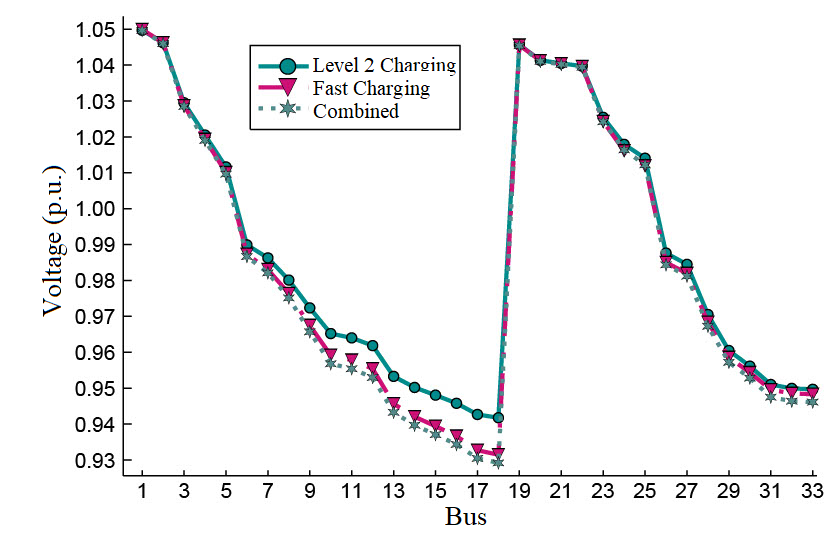}
    \caption{Voltage magnitude of buses at $6$ p.m.}
    \label{fig:time_18}
\end{figure}

Here, the voltage profile of buses and the power dispatch of DERs procured by the proposed method are presented. Note that the penetration level of EVs in this case is $50\%$. Besides, three scenarios are considered for this case study. In the "fast charging" scenario, all the EV charging points are equipped with fast-charging technology. In the "combined" scenario, some of the EV charging stations are equipped with fast-charging technology and the rest of them are second-level charging points. In the "level $2$ charging" scenario, all EV charging stations are second-level charging points. Note that in this configuration, when EVs are connected to the distribution network through level 1 chargers, their battery may not have enough energy to take their trip. Thus, level $1$ chargers are not considered in the presented setup. Besides, the maximum charging power of fast chargers is considered $100$ kW while the one for level $2$ chargers and level $1$ chargers considered $35$ kW and $2.4$ kW, respectively.  The increase in real power demand at buses leads to a decrease in the voltage magnitude of buses. Increasing the charging level of EVs increases the real power demand of some buses as it is shown in Fig. \ref{fig:power_time_18}. Fig. \ref{fig:power_time_18} shows that the real power demand of EVs with fast chargers is more than the real power demand of EVs with level $2$ chargers at buses $15-17$ at $6$ p.m. Thus, the voltage magnitude of those buses when fast chargers are utilized is less than the $0.95$ p.u. as it is shown in Fig. \ref{fig:time_18}. A similar situation occurs at buses $24-25$. However, the solar generation unit connected to bus $24$ supplies the extra real power demand of EV fast-charging point. Thus, the voltage magnitude of buses $24-25$ procured by the proposed method with fast-charging EVs is approximately the same as the one procured by the proposed method with level $2$ charging EVs as presented in Fig. \ref{fig:time_18}. 

\begin{figure}[h!]
    \centering
    \includegraphics[width=\columnwidth]{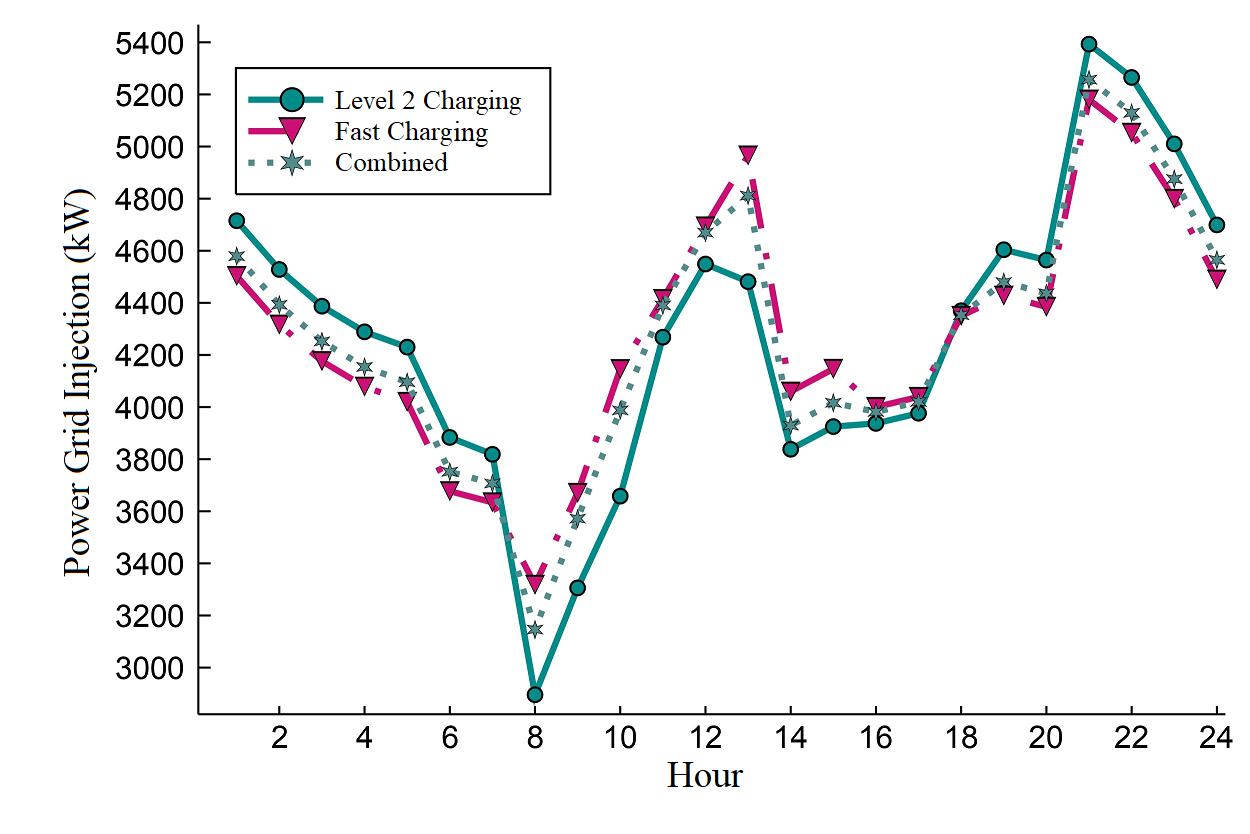}
    \caption{Main power grid injection to the distribution network over $24$ hours}
    \label{fig:pg}
\end{figure}

Charging EVs with higher levels increases the battery degradation cost. 
Utilizing fast-charging EVs enables EVs to charge when the price of electricity is lower given the time-of-use prices. The total cost of the system procured by the "fast charging" scenario is $\$3,157,157$, while the one procured by the "combined" and "level $2$ charging" scenarios are $\$3,174,215$ and $\$3,212,047$, respectively. Thus, utilizing fast chargers decreases the total operation cost of the system. Fast-charging EVs will enable the distribution network to decrease the real power dispatch from the grid during peak hours and buy electricity during off-peak hours to charge EVs as shown in Fig. \ref{fig:pg}.

\begin{figure}[h!]
\centering
    \includegraphics[width=\columnwidth]{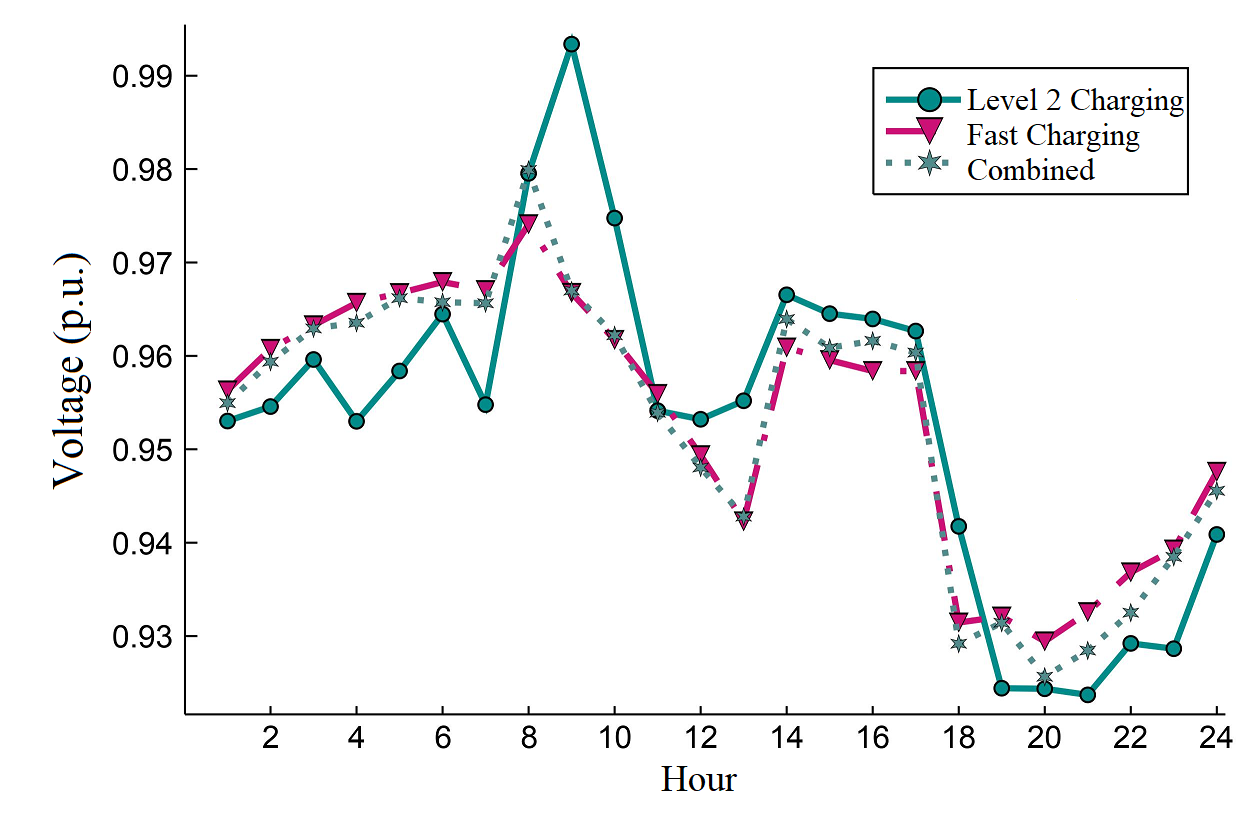}
    \caption{Comparing the impact of various charging scenarios on the voltage profile of bus $18$ over $24$ hours}
    \label{fig:bus-18}
\end{figure}

\begin{figure}[h!]
  \centering
  \includegraphics[width=\columnwidth]{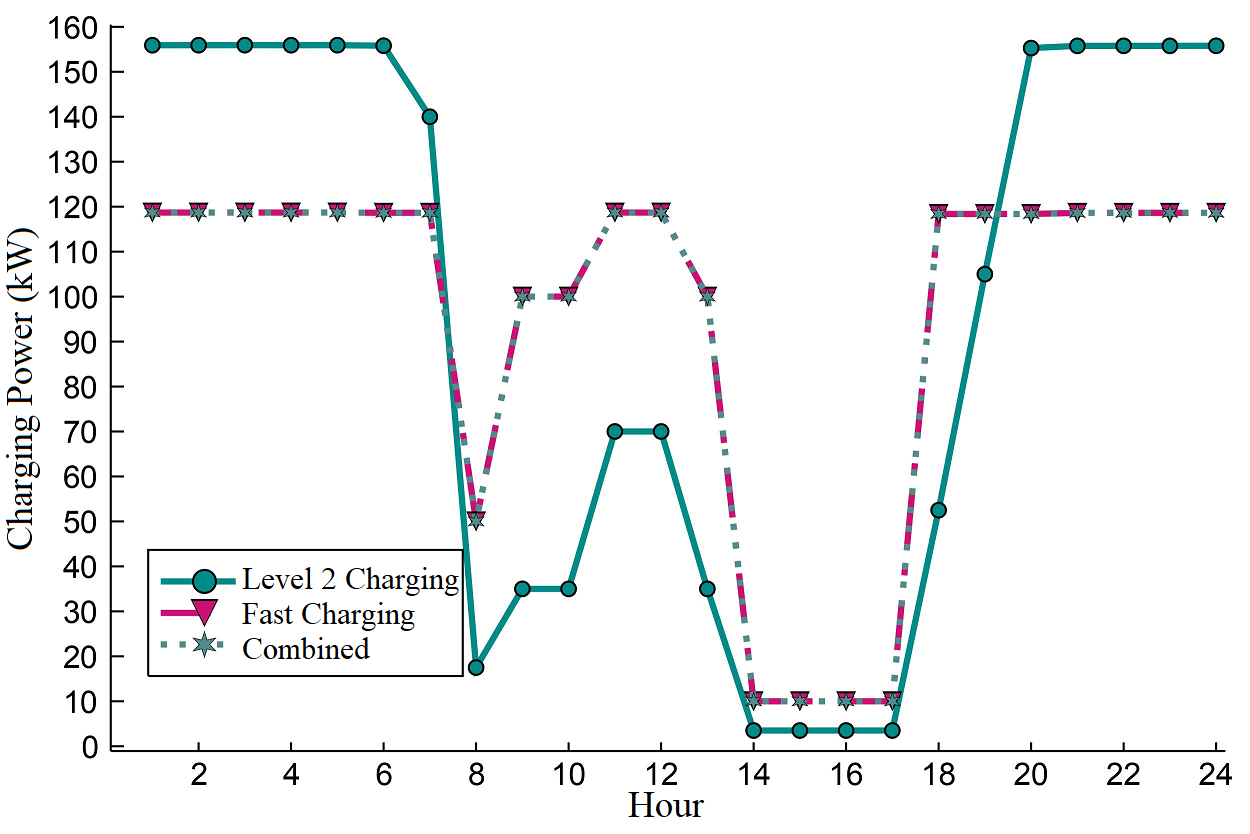}
  \caption{Comparing various scenario of the EV charging power at bus $18$ over $24$ hours}
  \label{fig:power-bus-18}

\end{figure}

In the combined scenario, the EV connected to bus $18$ is a fast-charging EV as presented in Fig. \ref{fig:power-bus-18}. Thus, the voltage profile of bus $18$ procured by the "fast charging" scenario is similar to the voltage profile of bus $18$ procured by the "combined" scenario as shown in Fig. \ref{fig:bus-18}. Voltage magnitude of bus $18$ is less than $0.95$ p.u. for $9$ hours in "fast charging" and "combined" scenarios, while the voltage magnitude of bus $18$ is less than $0.95$ p.u. for $7$ hours in "level $2$ charging" scenario as presented in Fig. \ref{fig:bus-18}. Thus, high-level EV charging will potentially cause voltage violations that should be treated by grid operators. 

\subsection{Impact of the Size of Solar Generation Units on the Voltage Profile of Buses}
\begin{figure}[h!]
    \centering
    \includegraphics[width=\columnwidth]{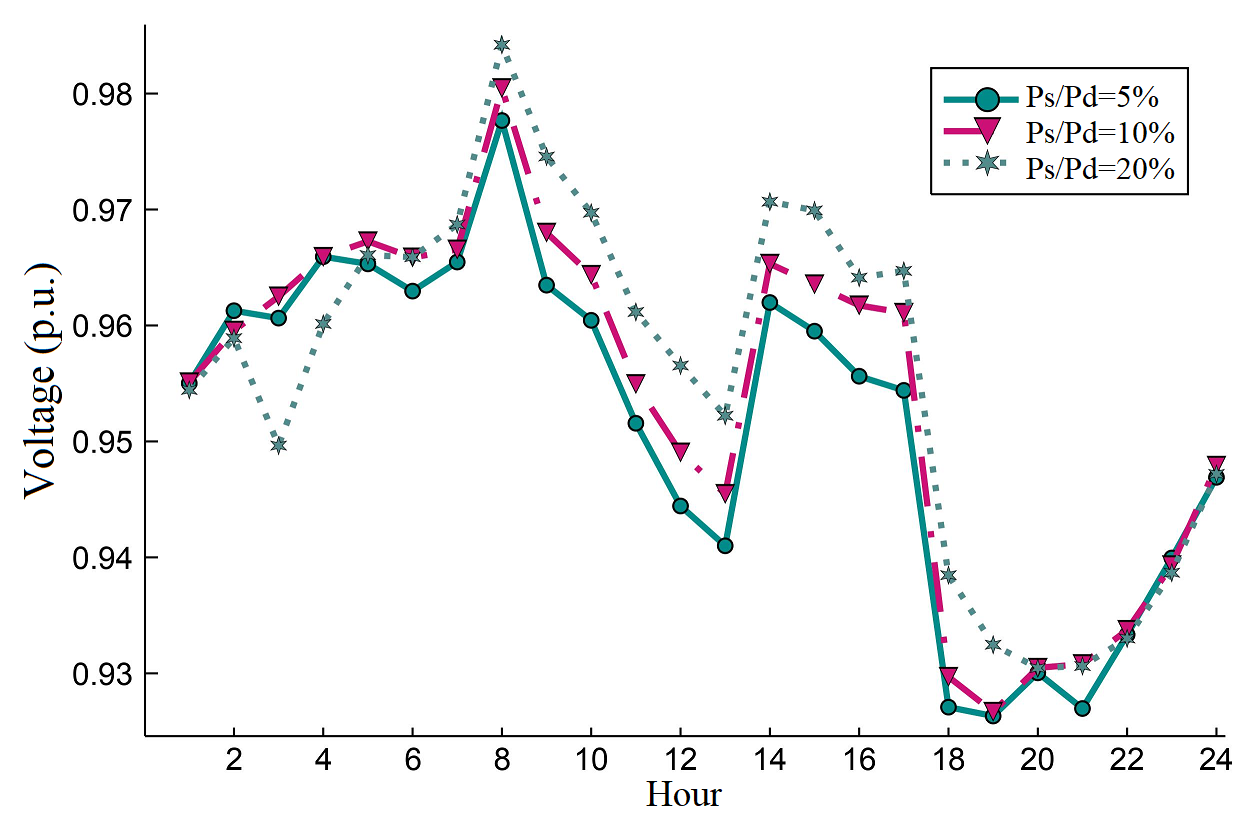}
    \caption{Voltage profile of bus $18$ over $24$ hours}
    \label{fig:bus_18_solar}
\end{figure}

\begin{figure}[h!]
    \centering
    \includegraphics[width=\columnwidth]{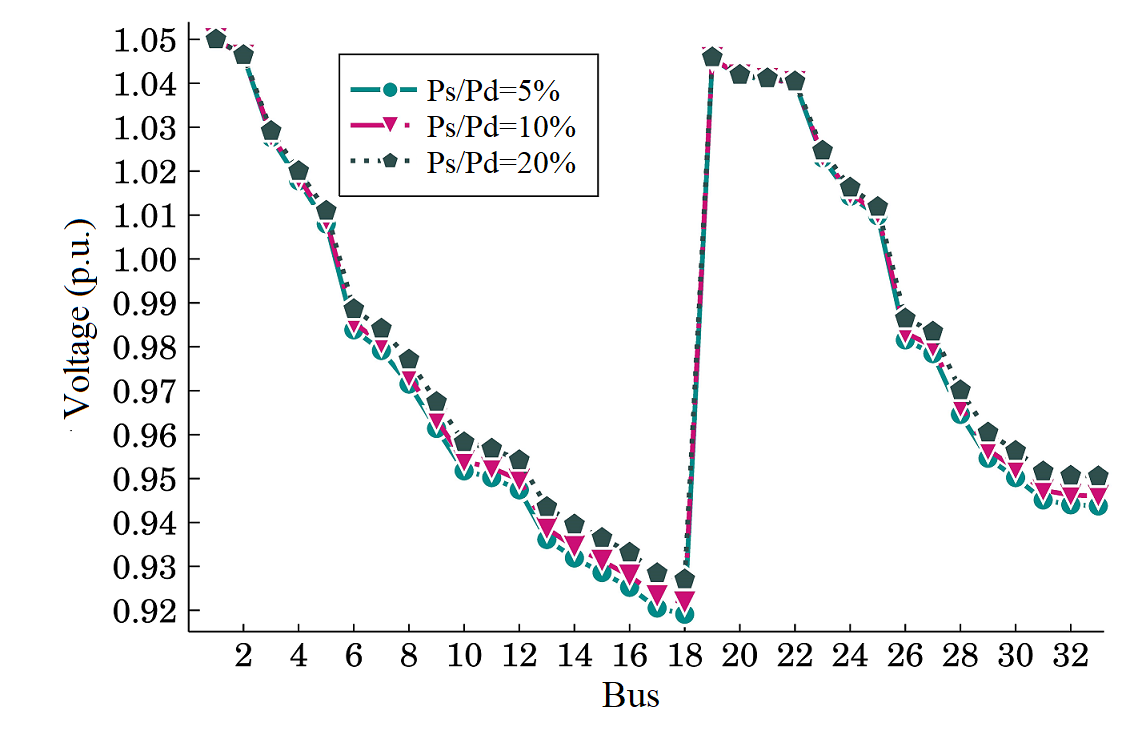}
    \caption{Voltage magnitude of buses at $6$ p.m. }
    \label{fig:time_18_solar}
\end{figure}

Here, the impact of available solar power on the voltage profile of buses is investigated. In this case, it is assumed that all EVs are fast-charging. Here, $3$ scenarios are considered. In the first scenario, the summation of available solar power divided by the summation of the real power of loads i.e. solar penetration level is $5\%$. While in the second and third scenarios, this ratio is $10\%$ and $20\%$, respectively. Increasing the available solar power generation decreases the voltage magnitude deviation of buses as presented in Figs. \ref{fig:bus_18_solar} and \ref{fig:time_18_solar} for bus $18$ over $24$ hours and for all buses at hour $18$, respectively.  

\begin{figure}[h!]
    \centering
    \includegraphics[width=\columnwidth]{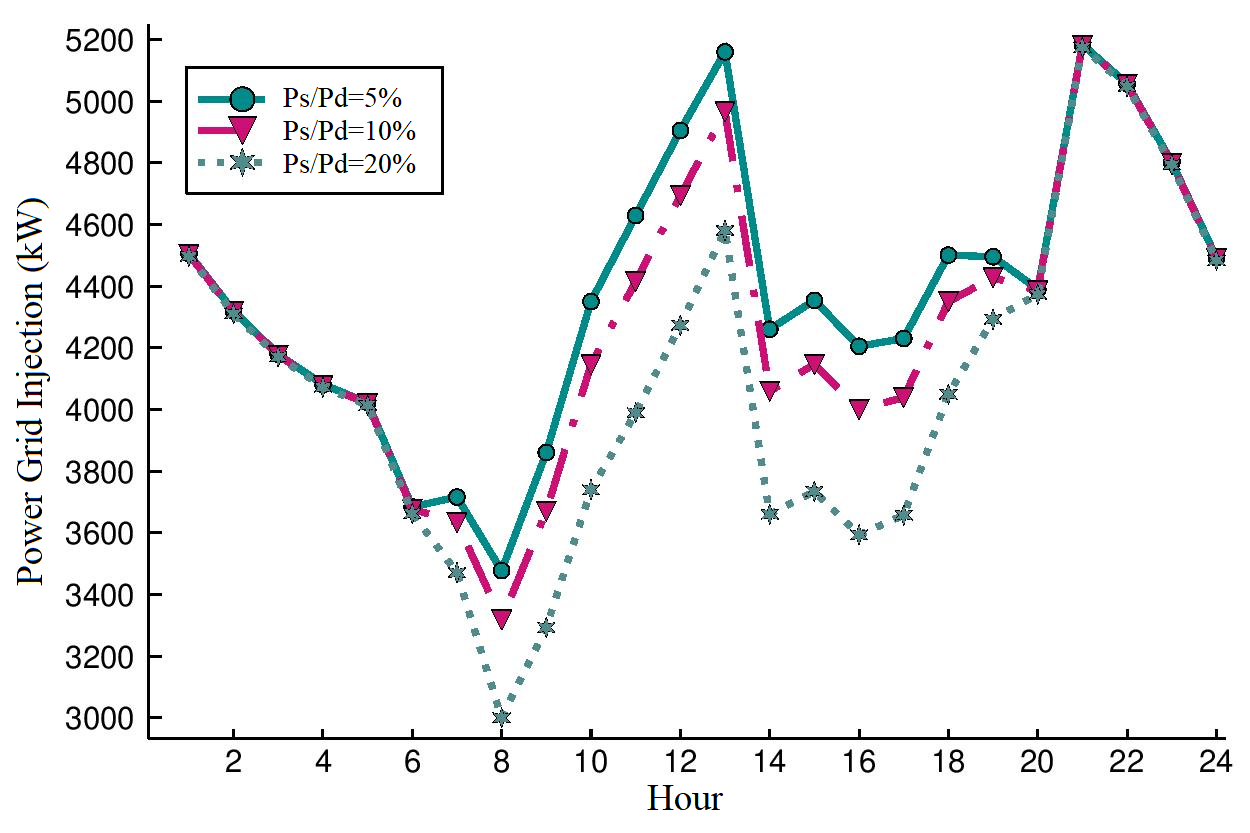}
    \caption{Main power grid injection to the distribution network over $24$ hours}
    \label{fig:pg_solar}
\end{figure}

An increase in the available solar power results in a decrease in the total operation cost of the system. This is as a result of a decrease in the real power dispatch of the grid to the distribution network in daylight hours as shown in Fig. \ref{fig:pg_solar}. The operation cost of the distribution network over $24$ hours procured by the proposed method under the first scenario is $\$ 3,223,414$, while those procured by the proposed method under scenarios $2$ and $3$ are $\$ 3,157,157$ and $\$ 3,025,381$, respectively.

\subsection{Impact of Degradation Cost on the Operation of Distribution Network}
\begin{figure}[h!]
    \centering
    \includegraphics[width=\columnwidth]{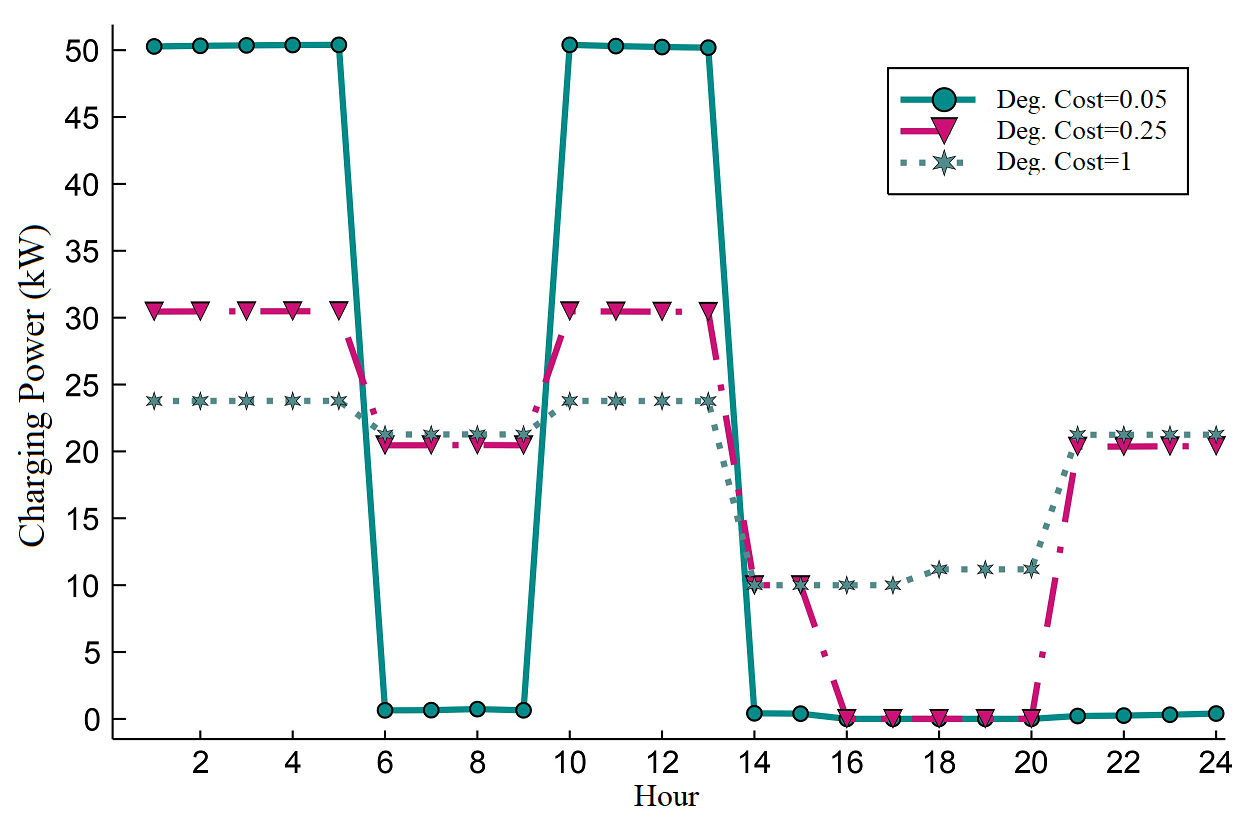}
    \caption{Charging of the EV connected to bus $18$ over $24$ hours}
    \label{fig:bus18_p_deg}
\end{figure}

\begin{figure}[h!]
    \centering
    \includegraphics[width=\columnwidth]{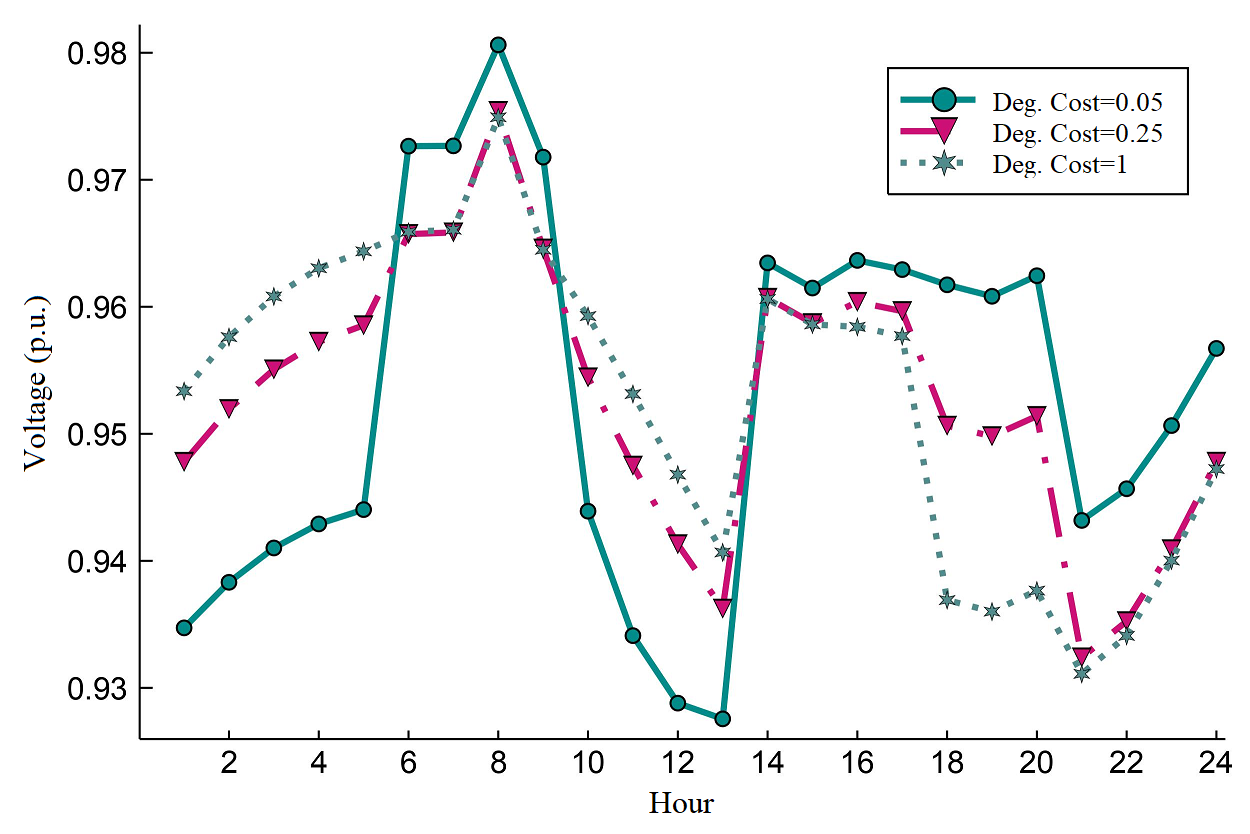}
    \caption{Voltage magnitude of bus $18$ over $24$ hours}
    \label{fig:bus18_deg}
\end{figure}

In this case, the impact of degradation cost of the batteries of EVs on the voltage profile of buses and the total operation cost of the distribution network is investigated. In this case, it is assumed that all EVs are fast-charging. Three scenarios are considered to illustrate the effect of changing battery degradation cost on the operation of the distribution network. In the first scenario, the degradation cost of EVs is $0.05$ $\$/100kWh$, while the degradation cost of the second and third scenarios are $0.25$ $\$/100kWh$ and $1$ $\$/100kWh$. Increasing the degradation cost while the time-of-use price is not changing will result in an increase in the charging power of EVs at peak hours as shown in Fig. \ref{fig:bus18_p_deg}. When the degradation cost is is very low in comparison with the time-of-use price, the EVs are charged when the price of electricity is lowest as presented in Fig. \ref{fig:bus18_p_deg} at hours $1-5$ and $10-13$. However, when the degradation cost increases to $1$ $\$/100kWh$, the EV tends to charge at a lower level. Thus, the charging level of EVs is decreased in off-peak hours and increased during peak hours. 
The violation of voltage magnitude of buses increases at off-peak hours as a result of charging EVs at off-peak hours when the degradation cost of batteries is low as shown in Fig. \ref{fig:bus18_deg}. When the degradation cost of the batteries increases, EVs will adopt a lower charging level. Thus, the voltage magnitude of buses increases during off-peak hours as presented in Fig. \ref{fig:bus18_deg}.

\section{Conclusion and Future Works}
In this paper, the impact of utilizing high penetration EVs with different charging levels on the power quality delivered by the distribution network is investigated. To this end, the SOCP relaxed form of the ACOPF of the distribution network in the presence of PV and EV charging stations is presented. 
The results illustrate that leveraging fast chargers exacerbates the voltage drop caused by utilizing high penetration EVs while decreasing the total operation cost of the system. However, it will cause violations in the voltage requirements across the distribution network. Besides, it is illustrated that an increase in the available solar power decreases the total operation cost of the electricity distribution network by decreasing the power grid injection to the distribution network and mitigates the voltage magnitude violation of buses. Results show that increasing the battery degradation cost mitigates the voltage magnitude violation of buses by shifting the charging power of EVs during $24$ hours. Investigating the impact of high penetration EVs with vehicle-to-grid capability on the power quality of the distribution system with different time-of-use pricing is an extension to this paper. 

\bibliographystyle{IEEEtran}

\bibliography{ref.bib}
\end{document}